\documentclass[reprint,
superscriptaddress,
longbibliography,
amsmath,amssymb,
aps,
prl,
floatfix,
]{revtex4-1}
\usepackage{graphicx}
\usepackage{dcolumn}
\usepackage{bm}
\usepackage[T1]{fontenc}
\usepackage[utf8]{inputenc}
\usepackage{color}
\usepackage[colorlinks = true,
            linkcolor = blue,
            urlcolor  = blue,
            citecolor = blue,
            anchorcolor = blue]{hyperref}
\usepackage{filecontents}

\begin{document}

\preprint{}

\title{Experimental Evidence of Non-Negligible Imaginary Part of Spin Mixing Conductance and Its Impact on Magnetization Dynamics in Heavy-Metal|Ferromagnet Bilayers}

\author{Adam Krysztofik}
\affiliation{Institute of Molecular Physics, Polish Academy of Sciences, ul.~Smoluchowskiego 17, 60-179 Pozna{\'{n}}, Poland}

\author{Piotr Graczyk}
\affiliation{Institute of Molecular Physics, Polish Academy of Sciences, ul.~Smoluchowskiego 17, 60-179 Pozna{\'{n}}, Poland}

\author{Hubert G{\l{}}owi{\'{n}}ski}
\affiliation{Institute of Molecular Physics, Polish Academy of Sciences, ul.~Smoluchowskiego 17, 60-179 Pozna{\'{n}}, Poland}

\author{Emerson Coy}
\affiliation{NanoBioMedical Centre, Adam Mickiewicz University, ul.~Wszechnicy Piastowskiej 3, 61-614 Pozna{\'{n}}, Poland}

\author{Karol Za{\l{}}{\k{e}}ski}
\affiliation{NanoBioMedical Centre, Adam Mickiewicz University, ul.~Wszechnicy Piastowskiej 3, 61-614 Pozna{\'{n}}, Poland}

\author{Iwona Go{\'{s}}cia{\'{n}}ska}
\affiliation{Institute of Molecular Physics, Polish Academy of Sciences, ul.~Smoluchowskiego 17, 60-179 Pozna{\'{n}}, Poland}

\author{Janusz Dubowik}
\email{dubowik@ifmpan.poznan.pl}
\affiliation{Institute of Molecular Physics, Polish Academy of Sciences, ul.~Smoluchowskiego 17, 60-179 Pozna{\'{n}}, Poland}

\date{\today}

\begin{abstract}

The paper concerns experimental verification of the magnitude of imaginary part of spin mixing conductance in bilayers comprising heavy metals.
We present  results of broadband ferromagnetic resonance studies on heterostructures consisting of Finemet thin films covered by Pt and Ta wedge layers with the aim to observe spin pumping effects and to evaluate both the real and imaginary parts of the spin mixing conductance.
The experimental results are analyzed in the framework of a recent microscopic theory which allows us to estimate the value of the interfacial spin-orbit interaction and confirm its important role.
In particular, we show that the imaginary part of spin mixing conductance cannot be regarded as negligible and we discuss its influence on magnetization dynamics.
For Finemet|Ta bilayers, the ratio  $\mathrm{Re}[g_{eff}^{\uparrow\downarrow}]/\mathrm{Im}[g_{eff}^{\uparrow\downarrow}]=$ 0.38, that is, the field-like torque dominates over the damping-like torque in such system.
Consequently, relatively small enhancement of precession damping with respect to significant total torque exerted on the magnetization in Finemet|Ta system offers an attractive perspective for an application in the next-generation magnetic random-access memory cells.

\end{abstract}

\maketitle

\section{Introduction}
\label{sI}

Increasing demand for data storage has intensively driven research efforts for new memory technologies in the past decades. Within discovery of the GMR effect, a concept of magnetoresistive RAM (MRAM) emerged, in which the resistance of two separated ferromagnetic films with parallel or antiparallel alignment of magnetization differentiated state "1" and state "0". A use of external magnetic field to switch the free layer however, hindered the scalability and required additional power for Oersted field generation \cite{b2}. A promising perspective of overcoming these barriers has been brought with a prediction of spin-transfer (ST) effect by Slonczewski and Berger \cite{Slonczewski1996,Berger1996}. Spin-polarized electron injection (from a pinned reference layer) was utilized in order to exert a ST torque on the free layer magnetization and, for  sufficient current strengths, to change its direction \cite{Garello2014}. Operation principle of STT-MRAM cell was now realized by local means and by incorporation of MgO layer as a spacer, high tunnel magnetoresistance accompanied by small switching current density were achieved \cite{Suess2017}. Notwithstanding these advantages, another technologic bottlenecks were recognized. The first one concerns the cell endurance delimited with aging of the thin tunnel barrier \cite{Sverdlov2018,Wang2018}. The second inconvenience was that reading and writing process was provided by a common current path that allows a possibility of false write operation during the bit readout \cite{Sverdlov2018,Cubukcu2018}. To decouple read/write current path yet another solution was proposed for next-generation MRAMs.

Instead of using spin-transfer torque, spin-orbit torque in a heavy-metal|ferromagnet bilayer is currently being envisioned as an effective mechanism for magnetization reversal \cite{Lee2018,Prenat2016,Bapna2018,Fan2019}. When the bilayer is implemented into cell architecture on the separate write line, the tunnel barrier reliability issues are eliminated and independent optimization process is feasible. Originating from the bulk spin Hall effect and/or the interfacial Rashba-Edelstein effect \cite{Bose2018,Wong2019}, the spin-orbit torque arises from non-equilibrium spin accumulation at the bilayer interface, and can be decomposed into two components: a (anti)damping torque and a field-like torque. Much scientific effort has been put to recognize key parameters characterizing heavy-metal|ferromagnet bilayers that ensure the most efficient switching \cite{Sverdlov2018,Gweon2019,Garello2013}. One of them is the spin Hall angle defined as conversion efficiency between charge currents and pure spin currents that lead to spin accumulation generation by heavy-metal film \cite{Zhu2018,Ramaswamy2017}. In terms of magnitude of spin current transferred into the ferromagnetic layer, equally important role plays a transparency of the bilayer interface which is described by the so-called spin-mixing conductance \cite{Zhang2015}. Moreover, for current-induced spin-orbit torques, it has been shown that the real part of the spin mixing conductance contributes to the damping-like torque while the imaginary part contributes to the field-like torque \cite{11}. The significance of the latter one is the subject of investigation of this paper.

Determination of spin mixing conductance is usually performed with the spin torque ferromagnetic resonance \cite{b1}. Alternative method involves the use of reciprocal effect that is spin pumping, commonly detected with vector network analyzer ferromagnetic resonance (VNA-FMR) measurements \cite{2,Wang_2018}. 
While the enhancement of magnetization precession damping due to spin pumping has been extensively studied enabling the evaluation of the real part of spin mixing conductance  \cite{3,4, 7,33,34,36,Wang2019,Yoon_2017,Lourembam2018}, the imaginary part has been only scarcely mentioned in literature since it is commonly regarded 1-2 order of magnitude smaller and therefore, negligible \cite{5,6,Zhu2018}. However, in the very first ferromagnetic resonance (FMR) experiment by Mizukami et al. \cite{7}, apparent changes in the gyromagnetic ratio (corresponding to $\mathrm{Im}[g_{eff}^{\uparrow\downarrow}]$) were observed experimentally without comprehensive explanation.

From among few reports discussing the changes of the resonance field (or a “field-like” torque) in the context of spin pumping effects, it is worth to list a few in which either a field-shift or $\mathrm{Im}[g_{eff}^{\uparrow\downarrow}]$ have been mentioned.
By analyzing  spin-orbit torques and spin pumping in NiFe|Pt bilayers, Nan et al. have estimated that $\mathrm{Re}[g_{eff}^{\uparrow\downarrow}]=2.2 \times 10^{15}$ cm$^{-2}$ is only 3.7 times larger than $\mathrm{Im}[g_{eff}^{\uparrow\downarrow}]= 0.6 \times 10^{15}$ cm$^{-2}$ \cite{8}.
They have also demonstrated that the spin-orbit damping-like and field-like torques scale with interfacial spin current transmission.
Sun et al. have found that a strong damping enhancement in YIG|Pt bilayers is accompanied by a substantial shift in the resonance field \cite{9}.
They concluded that the spin pumping effects may originate partially from ferromagnetic ordering due to the magnetic proximity effect in Pt atomic layers near the interface.
However, they have not linked the resonance field shift with $\mathrm{Im}[g_{eff}^{\uparrow\downarrow}]$.
An elegant interpretation of the imaginary part of the spin mixing conductance generated by FM insulators with exchange-coupled local moments at the interface to a NM metal has been recently proposed by Cahaya et al. \cite{10}.
They assumed that a coherent motion of the proximity RKKY spin density (for example in Pt) is locked to the precessing magnetization of the local moments leading to a renormalization of the effective magnetic field.
Moreover, the recently discovered spin-Hall magnetoresistance \cite{Nakayama2013} offered a unique possibility to measure $\mathrm{Im}[g_{eff}^{\uparrow\downarrow}]$ for the interface of a normal metal in contact with magnetic insulator by exposing it to out-of-plane magnetic fields. However, the applied model itself was based under assumption that  $\mathrm{Re}[g_{eff}^{\uparrow\downarrow}]/\mathrm{Im}[g_{eff}^{\uparrow\downarrow}] \gg 1$.  \cite{Kosub2018, Vlietstra2013}

It is also important to point out that the field-like torque become lately the subject of discussion since experimental results concerning spin currents suggested that it may play decisive role governing magnetization dynamics, although the values of $\mathrm{Im}[g_{eff}^{\uparrow\downarrow}]$ have not been determined from the requisite measurements \cite{12}. Furthermore, it was shown that the domain wall reflection at sample edges of such heterostructure relies on the field-like torque that is crucial in deterministic SOT switching mechanism \cite{Yoon_2017}. In this work, we study spin pumping and spin current transport in Finemet films covered by platinum (Pt) and tantalum (Ta) with a strong emphasis on the role of the imaginary part of spin mixing conductance.

\section{MICROSCOPIC ANALYSIS OF SPIN PUMPING}
\label{sMAoSP}
Recently, consistent analysis of spin pumping has been presented by Tatara and Mizukami  \cite{13} for both metallic and insulating ferromagnets.
Let us summarize briefly the main results of their analysis.
In the scattering approach, the spin current pumped by FMR results in modification of the Gilbert damping parameter $\alpha$ and the gyromagnetic ratio $\gamma_0$ of a lone FM film via spin mixing conductance \cite{14,15}:
\begin{equation}\label{eq:1}
\tilde{\alpha}=\alpha_0 + {\frac{a^3}{4\pi S d_F}}\mathrm{Re}[g_{eff}^{\uparrow\downarrow}]
\end{equation}
and
\begin{equation}\label{eq:2}
\tilde{\gamma}=\gamma_0(1-{\frac{a^3}{4\pi S d_F}}\mathrm{Im}[g_{eff}^{\uparrow\downarrow}])^{-1},
\end{equation}
where $a$ is the lattice constant, $S$ is the magnitude of the localized spin, $d_F$ is the thickness of the ferromagnet.
As $a^3/4\pi S=\hbar \gamma_0/4\pi M_S$ and $\gamma_0 = g_0 \mu_B/ \hbar$, where $M_S$ is the saturation magnetization, $\hbar$ is the reduced Planck constant and $\mu_B$ is the Bohr magneton, Eqs. \ref{eq:1} and \ref{eq:2} can be rewritten to
\begin{equation}\label{eq:3}
\delta\alpha=\tilde{\alpha}-\alpha_0={\frac{g_0 \mu_B}{4\pi M_S d_F}}\mathrm{Re}[g_{eff}^{\uparrow\downarrow}]
\end{equation}
and
\begin{equation}\label{eq:4}
\delta g/g_0=(\tilde{g}-g_0)/g_0\cong{\frac{g_0 \mu_B}{4\pi M_S d_F}}\mathrm{Im}[g_{eff}^{\uparrow\downarrow}].
\end{equation}
Hence,
\begin{equation}\label{eq:5}
{\frac{\delta\alpha}{\delta g/g_0}}\cong{\frac{\mathrm{Re}[g_{eff}^{\uparrow\downarrow}]}{\mathrm{Im}[g_{eff}^{\uparrow\downarrow}]}}
\end{equation}
and the ratio of $\mathrm{Re}[g_{eff}^{\uparrow\downarrow}]$ to $\mathrm{Im}[g_{eff}^{\uparrow\downarrow}]$ can be easily estimated from the spin-pumping experiments.

In the microscopic analysis, the magnetization dynamics is derived by evaluating the spin accumulation in a nonmagnetic metal as a result of interface hopping.
One of the main consequences of the analysis \cite{13} is an approximation of enhancement of the Gilbert damping parameter:
\begin{equation}\label{eq:6}
\delta\alpha=\eta{\frac{a}{d_F}}{\frac{1}{\epsilon_F^2}}\tilde{t}_{\uparrow}^0\tilde{t}_{\downarrow}^0
\end{equation}
where $\epsilon_F$ is the Fermi energy and $\eta$ is a dimensionless coefficient roughly equal to one (see supplementary materials \cite{supp}).
$\tilde{t}_{\uparrow}^0$ and $\tilde{t}_{\downarrow}^0$ are the so-called hopping amplitudes which correspond to the energy required for an electron to hop between orbitals at a ferromagnetic - non-magnetic interface in the absence of the spin-orbit interaction.
Similarly, the change in the g-factor due to spin pumping can be rewritten from Eq. (84) in Ref. \cite{13} to
\begin{equation}\label{eq:7}
{\frac{\delta g}{g_0}}=\eta{\frac{a}{d_F}}{\frac{1}{\epsilon_F^2}}\tilde{\gamma}_{xz}(\tilde{t}_{\uparrow}^0+\tilde{t}_{\downarrow}^0),
\end{equation}
where $\tilde{\gamma}_{xz}$ represents the interface spin-orbit interaction having a unit of energy.
Equation (\ref{eq:7}) suggests that a high value of $\delta g/g_0$ is expected if a strong interface spin-orbit interaction exists.
Therefore, the ratio $\delta\alpha/(\delta g/g_0)$, which is directly accessible experimentally equals to
\begin{equation}\label{eq:8}
{\frac{\delta\alpha}{\delta g/g_0}}={\frac{\tilde{t}_{\uparrow}^0\tilde{t}_{\downarrow}^0}{\tilde{\gamma}_{xz}(\tilde{t}_{\uparrow}^0+\tilde{t}_{\downarrow}^0)}}.
\end{equation}

\section{SAMPLES STRUCTURE AND METHODS}
\label{SSaM}
To verify and experimentally examine the main results of the quoted theory, we fabricated F|NM bilayers using Fe$_{66.5}$Cu$_1$Nb$_3$Si$_{13.5}$B$_6$Al$_7$ (Finemet, F) as the ferromagnetic layer and Pt, Ta as a nonmagnetic metal (NM).
Instead of commonly employed Permalloy films, we extensively used Finemet for the following reasons: ($i$) exceptional smoothness of Finemet surfaces \cite{16}; ($ii$) low Gilbert damping parameter, smaller in comparison to Permalloy \cite{Coy_2018,17,18}; ($iii$) relatively small inhomogeneous linewidth $\Delta H_0$ of $3-6$ Oe indicating a low density of defects \cite{16}.
The Finemet films with different thicknesses were deposited at room temperature by pulsed laser deposition (base pressure of $8\times 10^{-8}$ mbar) on naturally oxidized Si substrates and covered \textit{in situ} with wedge-shaped capping layers of Pt or Ta ($0 < d_{NM} < 7$ nm) using RF magnetron sputtering.
The nominal thicknesses of Finemet: $d_F = 10, 15, 20, 30, 40$ nm for the Pt cover layer and $2, 3.5, 5, 10$ nm for the Ta cover layer, were confirmed with X-ray reflectivity measurements, while thickness profiles of the NM capping layer were additionally verified using the energy dispersive spectroscopy in a scanning electron microscope.
Atomic force microscopy measurements of Finemet surface yielded RMS roughness of 0.1 nm.

\begin{figure}[]
 \includegraphics[width=1\linewidth]{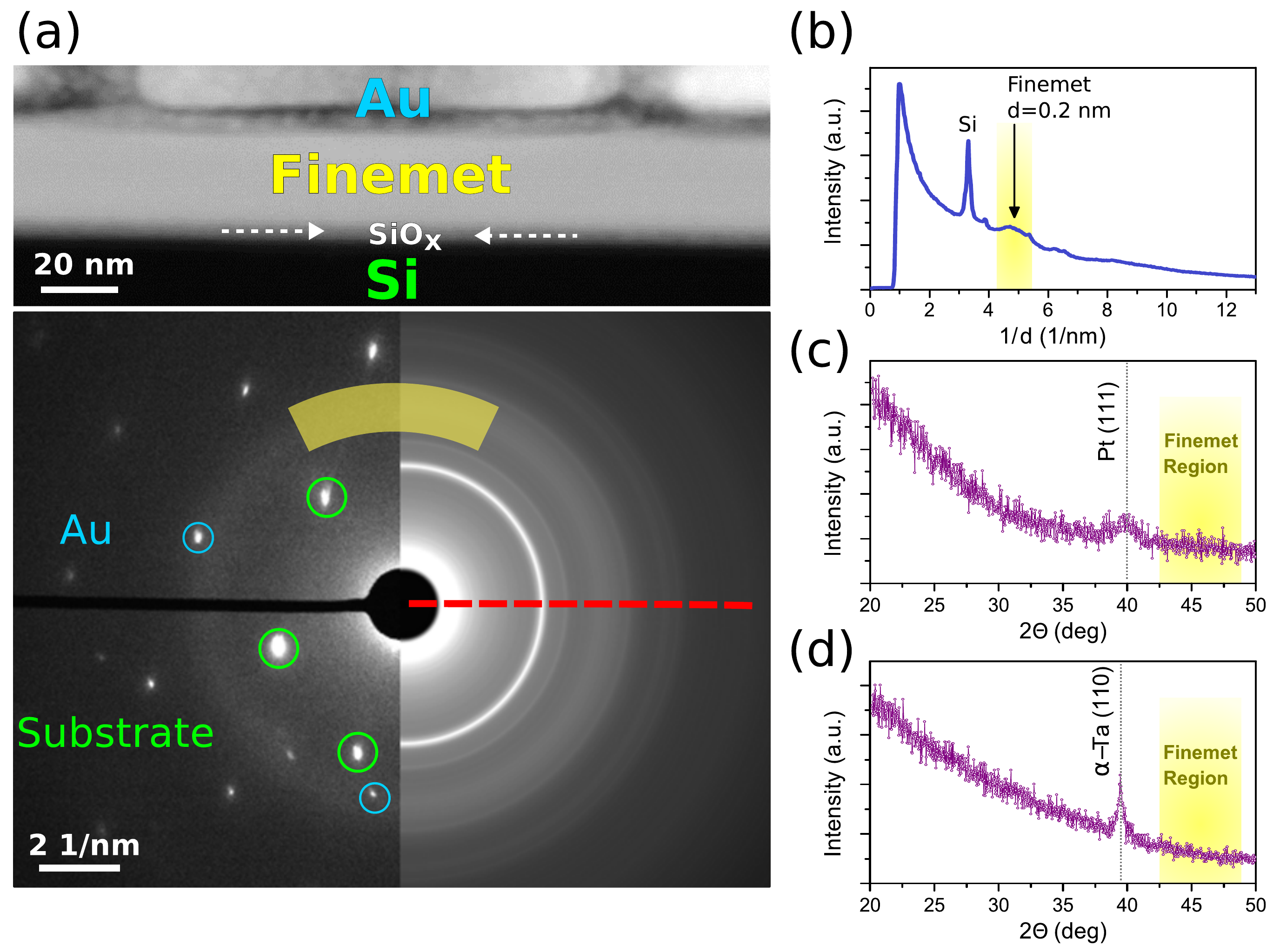}
\caption{(a) Low magnification scanning electron image of Finemet control layer (25 nm) coated ex-situ with Au for cross-section preparation, and the corresponding SAED pattern including both the substrate and the Au covering film (bottom panel).
The dashed red line shows the direction of the profile taken from the rotational averaging, displayed in (b).
The profile image shows the clear position of Finemet (d = 0.2 nm) plus the highly intense Si peaks.
Finally, (c) and (d) shows the GiXRD spectra of Finemet (30nm)|Pt (6nm) and Finemet (10nm)|Ta (6nm) respectively, where the texture of Pt(111) and (d) $\alpha$-Ta(110) covering layers can be seen.}
\label{fig:1}
\end{figure}

In order to determine the structural properties of the films, we conducted HR-TEM experiments and grazing incident X-ray diffraction.
As shown in an electron diffraction pattern (SAED) in Fig. \ref{fig:1}(a), a diffuse ring marked in yellow is observed, indicating the occurrence of a nanocrystalline phase in the Finemet film.
Apart from substrate peaks, reflections corresponding to a gold layer are visible, with which Finemet film was coated before the preparation of a cross-section by the focused ion beam.
The nanocrystalline band corresponds roughly to the interplanar distance of $0.2 \pm 0.1$ nm as presented in Fig. \ref{fig:1}(b), which was extracted by using a rotational average of the Digitalmicrograph\texttrademark, Diff tools plug-in \cite{19}.
This result is congruent with the simulation of Finemet unit cell (using the CaRIne Crystallography software, Lattice 2.8) yielding $d_{(110)} = 0.1980$ nm.
However, the highest intensity reflection (110), expected to be at 45.75 angle 2$\Theta$, is not observed in the grazing incident X-ray diffraction as marked by shaded areas in Fig. \ref{fig:1}(c) and (d).
Nevertheless, these measurements allowed us to determine the structure of capping Pt and Ta layers.
We conclude that the studied bilayer systems consist of nanocrystalline Finemet films covered with $\alpha$-Ta (i.e., bcc) or Pt with well-defined (110) and (111) textures, respectively.

\begin{figure*}[]
 \includegraphics[width=0.7\linewidth]{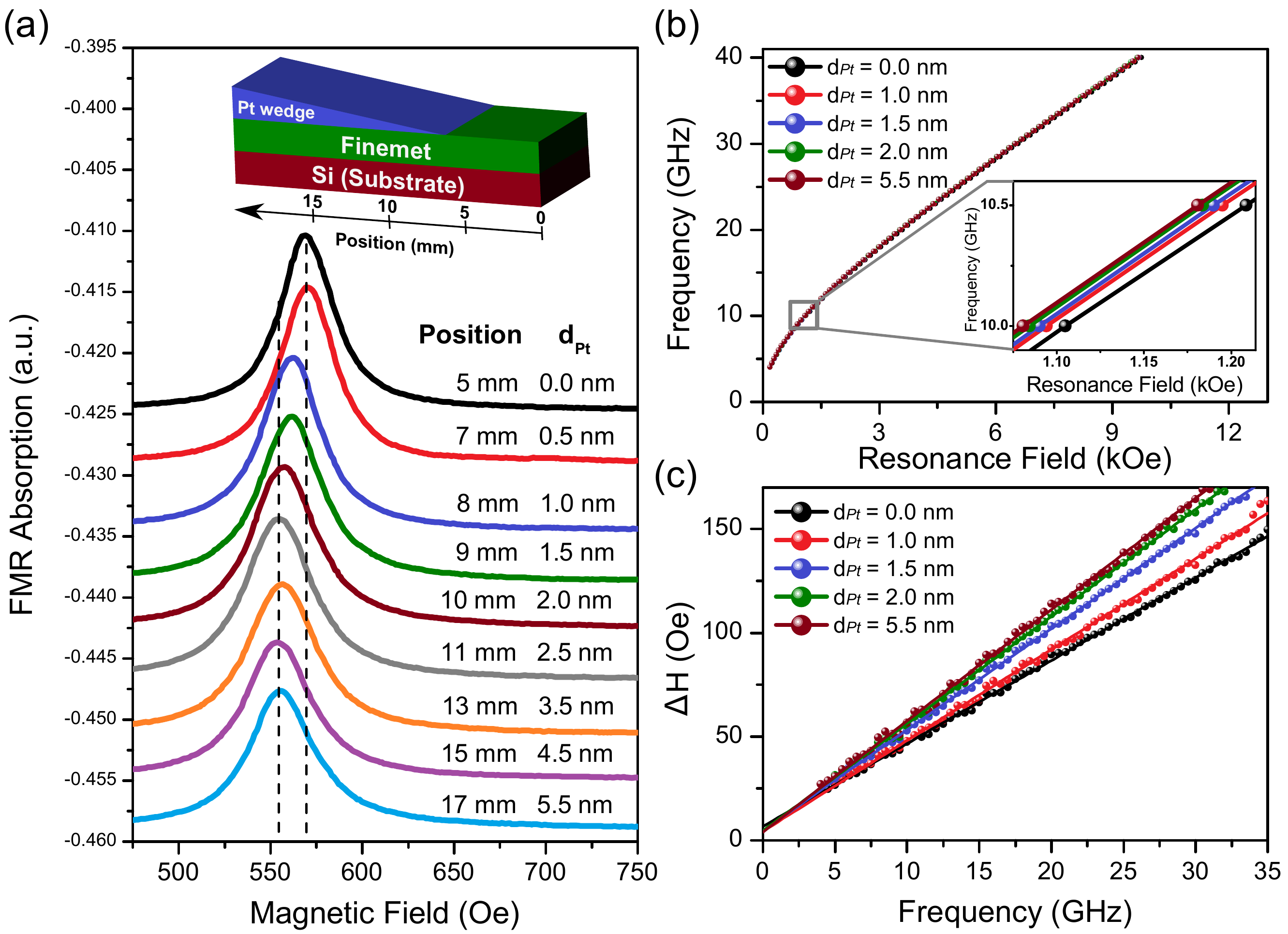}
\caption{VNA-FMR results for a 30 nm thick Finemet film covered with Pt.
(a) Typical FMR spectra taken at $f = 7$ GHz for various positions along the wedge-shaped Pt layer.
(b) Frequency vs resonance field dependences fitted using Eq. \ref{eq:9}.
The inset shows enlarged region near $f = 10$ GHz.
(c) Linewidth dependences on frequency fitted using Eq. \ref{eq:10}.}
\label{fig:2}
\end{figure*}

Ferromagnetic resonance measurements of the samples were carried out at room temperature on a coplanar waveguide (CPW) in the in-plane configuration over a frequency range of 4 - 40 GHz, as detailed in Ref. \cite{20,21}.
For each thickness $d_{NM}$ we measured the field swept complex transmission parameter S$_{21}$(H) by placing the sample at a certain position along the wedge (Fig. \ref{fig:2}(a)).
The real and imaginary transmission spectra were fitted using the Lorenzian and anti-Lorentzian function in a similar way as described in Ref. \cite{22}.
From the fits, the resonance field $H_r$ and resonance linewidth $\Delta H$ (full width at half maximum) were evaluated at several frequencies $f$.
In short, our method involves the use of a single sample with a fixed thickness $d_F$ wherein the capping layer is wedge shaped, so by scanning the sample over the central line of CPW one can determine parameters of magnetization dynamics as a function of $d_{NM}$.
Such an approach is of crucial importance, because it eliminates the problem of the proper choice of a reference layer \cite{23}.
Since deposition conditions (and consequently film properties) may slightly differ in each sputtering or ablation process we have set the beginning of the NM wedge at 6-7 mm from the substrate’s edge allowing for the measurements of a lone Finemet film in every F|NM bilayer (see the inset of Fig. \ref{fig:2}(a)).
To theoretically model the magnetization dynamics, we followed the methodology presented in Refs. \cite{24,25,26,27,28}.
Using finite element method in Comsol Multiphysics, the Landau-Lifshitz equation with spin torque term was numerically solved in the framework of the diffusive model.
Detailed information on the utilized formalism can be found in the supplementary material  \cite{supp}.
\section{EXPERIMENTAL RESULTS}
\label{ER}
Typical resonance spectra of Finemet (30 nm)|Pt (0 - 7 nm) sample, measured at $f = 7$ GHz for various positions along the wedge, are displayed in Fig. \ref{fig:2}(a).
It can be seen that an increase in linewidth is also accompanied by a substantial, yet unexpected shift in the resonance field $H_r$ as the Pt thickness increases.
The shift is negative (i.e., towards lower fields) and indicates systematical changes in the gyromagnetic ratio.
To obtain values of the $g$-factor we performed fittings to $f$ vs. $H_r$ dependences rather than evaluate changes in $H_r$ measured at one fixed frequency in order to not make an additional assumption of the effective magnetization value $M_{eff}$.
As it is shown in Fig. \ref{fig:2}(b), the relation follows Kittel’s dispersion $f = \frac{g \mu_B}{h} \sqrt{H_r(H_r + 4 \pi M_{eff})}$, however detailed analysis of the fitting residuals revealed that Finemet films possess a non-negligible anisotropy field $H_a$ (see supplementary materials \cite{supp}).
Hence, the model was extended to the following expression:
\begin{equation}\label{eq:9}
f = \frac{g \mu_B}{h} \sqrt{(H_r + H_a)(H_r + H_a + 4 \pi M_{eff})}.
\end{equation}
It should be emphasized that the inclusion of $H_a$ in Eq. \ref{eq:9} (along with $g$-factor and $M_{eff}$) results in three, coupled adjustable parameters that are not orthogonal to each other during the least-squares non-linear fitting process \cite{29}.
Therefore, we followed the methodology presented in Ref. \cite{29} and the applied asymptotic analysis to the data obtained over a finite range of frequencies to precisely determine the value of the $g$-factor as presented below.
Across the set of samples, we found that the Finemet films are characterized by $M_{eff}$ in the range of $750 \pm 30$ emu/cm$^3$ and $H_a$ in the range of $7 \pm 2$ Oe.

\begin{figure*}[t]
 \includegraphics[width=0.7\linewidth]{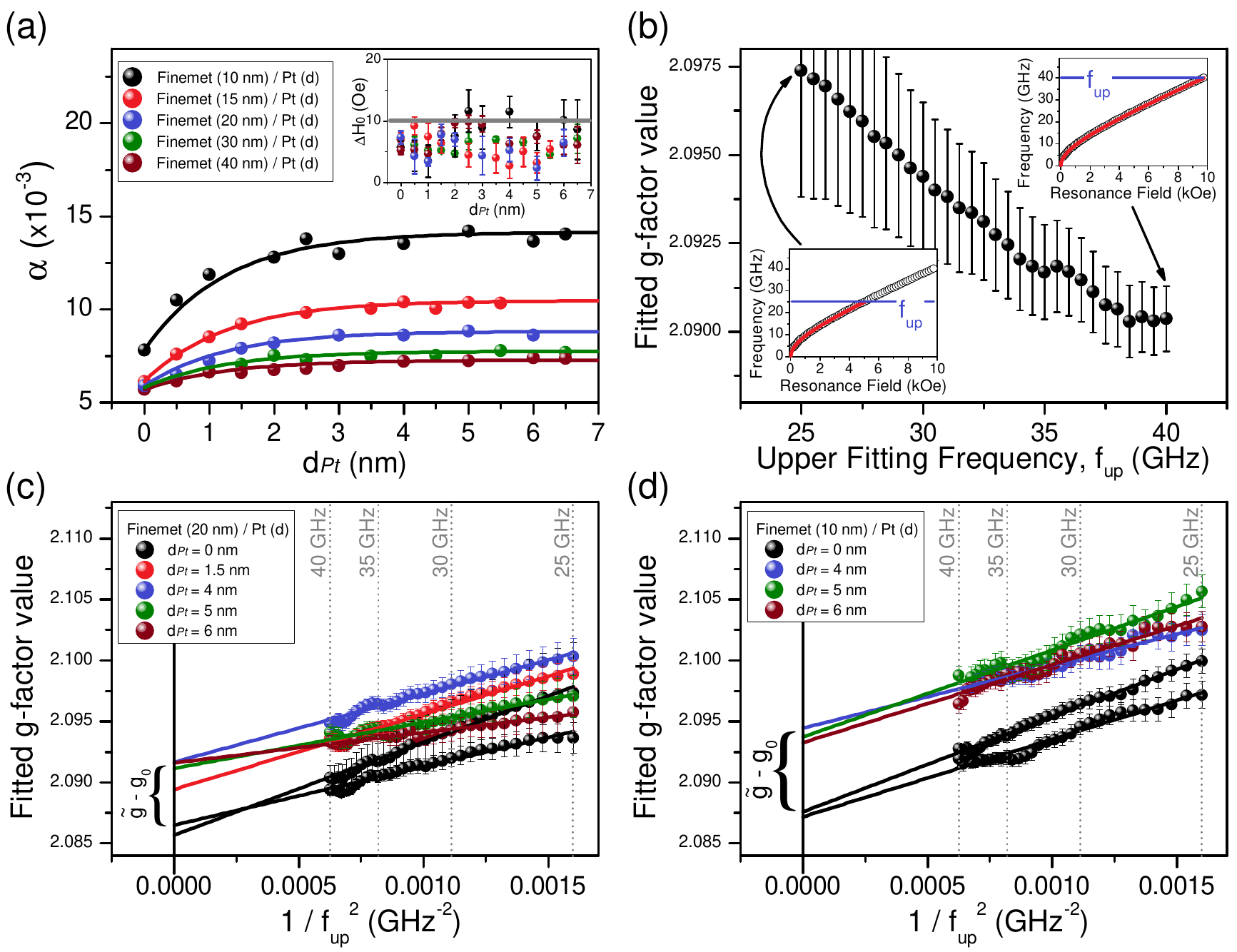}
\caption{(a) The Gilbert damping parameter $\alpha$ as a function of Pt film thickness.
The solid lines were obtained by simultaneous fit to all collected data using Eq. \ref{eq:11}.
The inset shows values of $\Delta H_0$ parameter.
(b) $g$-factor values of 20 nm thick Finemet fitted using Eq. \ref{eq:9} for different frequency ranges defined by $f_{up}$ as shown in the insets.
In (c) and (d), fitted values of the $g$-factor are plotted as a function of $f_{up}^{-2}$ respectively for 20 nm and 10 nm thick Finemet with different Pt thicknesses $d_{Pt}$.
Double sets of data for $d_{Pt} = 0$ nm (black symbols and lines) derive from measurements of the lone Finemet film taken at different positions outside the Pt wedge.}
\label{fig:3}
\end{figure*}

The Gilbert damping parameter $\alpha$ and the inhomogeneous broadening $\Delta H_0$ for a given $d_F$ and different thicknesses $d_{NM}$ were obtained from linear fits to the resonance linewidth $\Delta H$ as a function of frequency, according to the standard expression \cite{30}:
\begin{equation}\label{eq:10}
\Delta H(f) = \frac{4 \pi \alpha}{\gamma} f + \Delta H_0.
\end{equation}
Even for relatively thick Finemet films (for example 30 nm as in Fig. \ref{fig:1}(c)) it is clearly seen that the slopes of $\Delta H(f)$ dependences experience a substantial change with increasing thickness of Pt.
Indeed, as summarized in Fig. \ref{fig:3}, a continuous increase in the damping parameter $\alpha$ is observed for each investigated bilayer.

The damping of a lone Finemet is equal to $\alpha = (5.8 \pm 0.3) \times 10^{-3}$, though for the thinnest sample (10 nm) it is slightly elevated to the value of $\alpha = (7.80 \pm 0.07) \times 10^{-3}$ \cite{31}.
The $\Delta H_0$ parameter remains below 10 Oe in the investigated set with the mean value of 6.4 Oe indicating a low density of surface defects and, therefore, negligible extrinsic contributions to the linewidth such as two-magnon scattering (inset in Fig. \ref{fig:2}(a)).
Following the equation describing the damping enhancement due to spin pumping \cite{32}:
\begin{equation}\label{eq:11}
\alpha = \alpha_0 + \frac{g \mu_B}{4 \pi M_S} \frac{\mathrm{Re}[g_{eff}^{\uparrow \downarrow}]}{d_F}(1 - exp(-\frac{2 d_{NM}}{\lambda_{sf}})),
\end{equation}
the real part of spin mixing conductance $\mathrm{Re}[g_{eff}^{\uparrow \downarrow}] = (3.14 \pm 0.23) \times 10^{15}$  cm$^{-2}$, and additionally, spin diffusion length $\lambda_{sf} = 2.8 \pm 0.5$ nm were obtained.

\begin{figure}[]
 \includegraphics[width=0.9\linewidth]{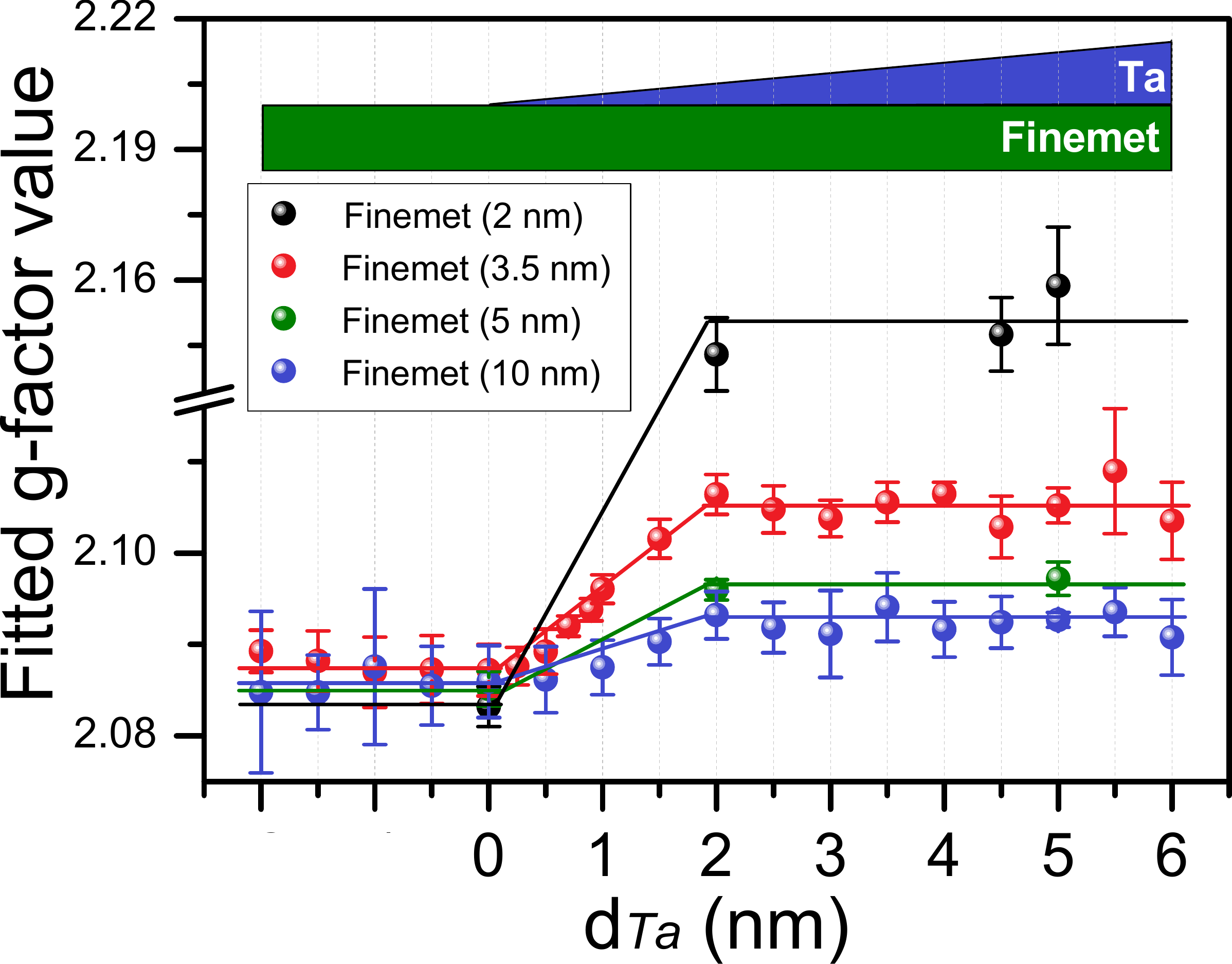}
\caption{Dependence of the $g$-factor on Ta layer thickness for Finemet|Ta bilayers with $d_F = 2 - 10$ nm.
Continuous lines serve as guides to the eye.}
\label{fig:4}
\end{figure}

In order to determine the imaginary part of $g_{eff}^{\uparrow \downarrow}$, an analysis of the $g$-factor was performed according to Eq. \ref{fig:4}.
First, the influence of a finite fitting range on the obtained value of the g-factor was examined.
As it is shown in Fig. \ref{fig:3}(b), with the increasing fitting range specified by the upper frequency $f_{up}$, the statistical error of the $g$-factor decreases and the value begin to saturate when plotted as a function of $f_{up}$.
It is considered that for large enough resonance fields, the fitted $g$-factor value becomes independent of a fitting range and therefore can be extrapolated from $g_{fit}(1/f_{up}^2)$ dependence \cite{29}.
Although this claim is not verifiable since every FMR spectrometer is limited either by available fields or frequencies, it should be highlighted that in terms of the analysis provided hereby, the method allows to reasonably determine the difference between $g$-factor values of the uncovered Finemet films and those capped with the non-magnetic metal.
As it can be seen in Fig. \ref{fig:3}(c) and (d), the difference $\delta g = \tilde{g} - g_0$ is essentially equivalent whether it is evaluated for $f_{up} = 40$ GHz or when $f_{up} \rightarrow \infty$ $(1/f_{up}^2 \rightarrow 0)$.
However, by taking the asymptotic values of $\tilde{g}$ and $g_0$ into account, one provides better statistical accuracy for such subtle changes in the $g$-factor.
For the Platinum thickness equal to 1.5 nm (depicted by a red line in Fig. \ref{fig:3}(c)), it even allows for an observation of intermediate value of the $g$-factor which falls between $\tilde{g}$ and $g_0$ and can be hardly concluded when the fitting is performed merely up to 40 GHz.
This suggests that with the appearance of nonmagnetic layer, the enhancement in the $g$-factor does not occur in an immediate, step-like way but follows a gradual increase until saturated.
As it is presented in Figs. \ref{fig:3}(c) and (d), asymptotically fitted $g$-factor for $d_{Pt} > 4$ nm already converge to nearly the same value $\tilde{g}$.
Moreover, the obtained difference $\delta g$ is greater for the thinner Finemet film, congruently with the theoretical prediction given by Eq. \ref{eq:4}.
For the 20 nm thick Finemet $\delta g = (5.1 \pm 0.4) \times 10^{-3}$, while for 10 nm thick layer $\delta g = (6.7 \pm 0.4) \times 10^{-3}$.
Such a result provides clear evidence of a spin pumping influence on magnetization dynamics expressed by the imaginary part of spin mixing conductance.
Following this observation, it is expected that the difference $\delta g$ will be larger for even thinner films.
For the set of samples with Ta deposited on 2-10 nm thick Finemet this influence is indeed much more pronounced as displayed in Fig. \ref{fig:4}.
Also, a quasi-linear enhancement in the $g$-factor is observed as the thickness of Ta increases from 0 up to 2 nm.

\begin{figure}[]
 \includegraphics[width=1\linewidth]{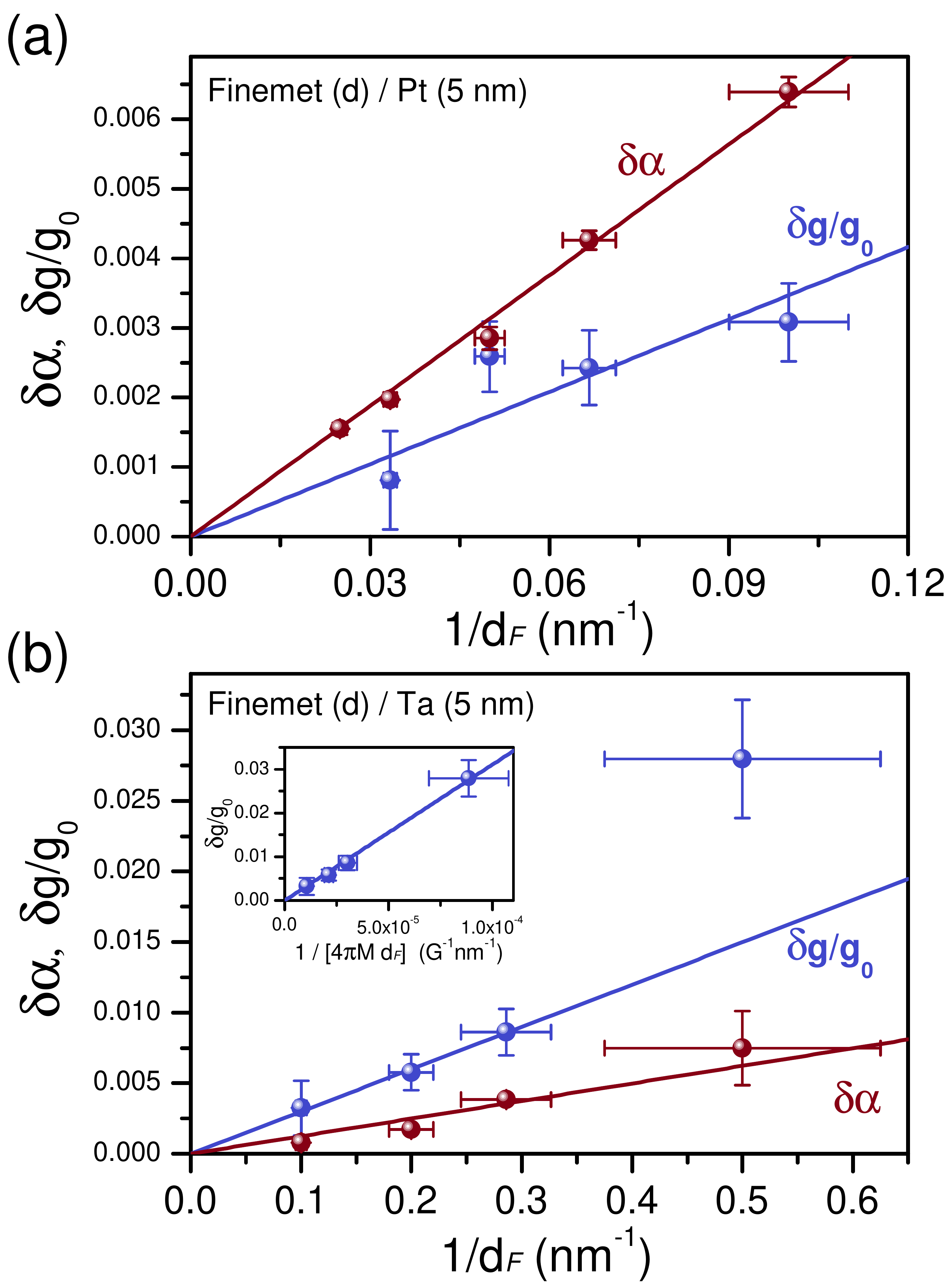}
\caption{Gilbert damping enhancement and relative change in the $g$-factor due to spin pumping for the Finemet|Pt (a) and Finemet|Ta bilayers (b).
Solid lines are fits to the data according to Eqs. \ref{eq:3} and \ref{eq:4}.
The inset in (b) shows that $\delta g/g_0$ scales almost perfectly with $(4 \pi M d_F)^{-1}$ if the magnetization of the film experiences serious changes with $d_F$ (see Eq. \ref{eq:4}).}
\label{fig:5}
\end{figure}

The main, experimental results of our paper are juxtaposed in Fig. \ref{fig:5}, where $\delta\alpha$ and $\delta g/g_0$ are plotted versus the inverse thickness $d_F$ for Finemet|Pt and Finemet|Ta bilayers.
The real and imaginary part of spin mixing conductance were determined from the slopes of linear fits according to Eqs. \ref{eq:3} and \ref{eq:4}.
For the sample covered with Platinum, $\mathrm{Re}[g_{eff}^{\uparrow\downarrow}] = (3.05 \pm 0.14) \times 10^{15}$ cm$^{-2}$ in agreement with the value obtained above from simultaneous fit to the data presented in Fig. \ref{fig:3}(a).
The $\mathrm{Im}[g_{eff}^{\uparrow\downarrow}]$ equals to $(1.69 \times 0.22) \times 10^{15}$ cm$^{-2}$, therefore the ratio $\mathrm{Re}[g_{eff}^{\uparrow\downarrow}]/\mathrm{Im}[g_{eff}^{\uparrow\downarrow}] = 1.81$ implies that the imaginary part of spin mixing conductance cannot be in general regarded as negligible in contrast to many common views \cite{b1,5,13}.
Strikingly different relation is observed for Finemet film covered by Tantalum.
It is clearly seen that the slope of $\delta g/g_0$ is substantially higher than the slope of $\delta\alpha$ vs $d_F^{-1}$.
Here, $\mathrm{Re}[g_{eff}^{\uparrow\downarrow}] = (0.61 \times 0.05) \times 10^{15}$ cm$^{-2}$ and $\mathrm{Im}[g_{eff}^{\uparrow\downarrow}] = (1.61 \times 0.07) \times 10^{15}$ cm$^{-2}$, hence the ratio $\mathrm{Re}[g_{eff}^{\uparrow\downarrow}]/\mathrm{Im}[g_{eff}^{\uparrow\downarrow}] = 0.38$.
It should be emphasized at this point that for the 2 nm-thick Finemet film ($1/d_F = 0.5$ nm$^{-1}$) the value of $\delta g/g_0$ significantly deviates from the linear relationship as can be seen in Fig. \ref{fig:5}(b).
We found that in such thin film the saturation magnetization was decreased down to $450 \pm 34$ emu/cm$^3$ resulting in the augmented value of $\delta g/g_0$ (see the Eq. \ref{eq:4}).
Therefore, in order to determine $\mathrm{Im}[g_{eff}^{\uparrow\downarrow}]$, the fitting was performed to $\delta g/g_0$ vs $(4 \pi M d_F)^{-1}$ dependence as shown in the inset of Fig. \ref{fig:5}(b).

\section{DISCUSSION}
\label{DN}
\subsection{Comparison of experimental results with other studies}
\label{CoERwOS}
For Finemet|Pt bilayers our results nearly agree with those obtained by Mizukami et al. \cite{7}.
Keeping in mind that they had two interfaces (Pt|Permalloy|Pt), their study yields $\mathrm{Re}[g_{eff}^{\uparrow\downarrow}]$ and $\mathrm{Im}[g_{eff}^{\uparrow\downarrow}]$ equal to $3 \times 10^{15}$ cm$^{-2}$ and $0.6 \times 10^{15}$ cm$^{-2}$, respectively.
For Finemet|Ta bilayers the impact of the Ta layer on damping is very reduced in comparison to Pt, in agreement with former studies \cite{7,33,34}.
The estimated real part of spin mixing conductance yielding $\mathrm{Re}[g_{eff}^{\uparrow\downarrow}]$ of $0.61 \times 10^{15}$ cm$^{-2}$, coincides with earlier reports for Permalloy|Ta ($0.51 \times 10^{15}$  cm$^{-2}$) \cite{7}, Co$_2$MnGe|Ta ($0.55 \times 10^{15}$ cm$^{-2}$) \cite{33} or for CoFeB|Ta ($0.54-1 \times 10^{15}$ cm$^{-2}$) \cite{34,Glowinski_2017} confirming that the weak damping enhancement for Ta arises from a small value of $\mathrm{Re}[g_{eff}^{\uparrow\downarrow}]$ \cite{35,36}.
Therefore, F|Ta shows an interesting property of a minor impact of spin pumping on $\alpha$, in accordance with Ref. \cite{23}.

In contrast, the results for the imaginary part of spin mixing conductance for Finemet|Ta bilayers show different behavior than that observed by Mizukami et al. for Permalloy|Ta structures \cite{7}.
The impact of Ta on the $g$-factor results in a rather high value of $\mathrm{Im}[g_{eff}^{\uparrow\downarrow}] = 1.61 \times 10^{15}$ cm$^{-2}$.
Their observations concerning the $g$-factor versus d$_{\text{Py}}$ are not confirmed by our measurements.
Instead of a substantial down shift in the $g$-factor corresponding to $\mathrm{Im}[g_{eff}^{\uparrow\downarrow}] = -0.46 \times 10^{15}$ cm$^{-2}$ (inferred from only one experimental point for thin Permalloy of 3 nm - see Fig. 6 in Ref. \cite{7}), we see in Fig. \ref{fig:5} a regular behavior that can be nicely fitted by using Eq. \ref{eq:4}.
Figure \ref{fig:4} further confirms that the increase in the $g$-value for F|Ta bilayers is regular and saturates for $d_{\text{Ta}} > 2$ nm.
As we show further on, the impact of a considerable value of $\mathrm{Im}[g_{eff}^{\uparrow\downarrow}]$ on spin transport in Ta may have important consequences in the magnetization dynamics.

For Finemet|Pt bilayers $\frac{\delta\alpha}{\delta g/g_0}$ (or equivalently, $\mathrm{Re}[g_{eff}^{\uparrow\downarrow}]$/$\mathrm{Im}[g_{eff}^{\uparrow\downarrow}]$) equals to 1.81, which is in an agreement to the order of magnitude with ratios one can infer from the results of Mizukami et al. ($\approx 5$) \cite{7} or Nan et al. ($\approx 3.8$) for Permalloy|Pt \cite{8}.
In contrast, a relatively high value of $\mathrm{Im}[g_{eff}^{\uparrow\downarrow}]$ for Finemet|Ta bilayers leads to $\frac{\delta\alpha}{\delta g/g_0} = 0.38$.
According to the microscopic theory, the difference between those two ratios originates from disparate values of the hopping amplitudes or spin-orbit interaction (see Eq. \ref{eq:8}).

\subsection{Interface spin-orbit interaction and hopping amplitudes}
\label{ISOIaHA}

By combining Eqs. \ref{eq:6} and \ref{eq:7} (see derivation in the supplementary materials \cite{supp}) we arrive at a simple quadratic equation with respect to the hopping amplitudes $\tilde{t}_{\sigma}^0$:
\begin{equation}\label{eq:12}
\frac{m_e^2 a^5}{2 \pi^2 d_F \hbar^4}(\tilde{t}_{\sigma}^0)^2 - \frac{\delta g/g_0}{\tilde{\gamma}_{zx}}\tilde{t}_{\sigma}^0 + \delta\alpha = 0,
\end{equation}
which is equivalent to:
\begin{equation}\label{eq:13}
\frac{m_e^2 a^5}{2 \pi^2 \hbar^4}(\tilde{t}_{\sigma}^0)^2 - \frac{\mathrm{Im}[g_{eff}^{\uparrow\downarrow}]}{\tilde{\gamma}_{zx}}\tilde{t}_{\sigma}^0 + \mathrm{Re}[g_{eff}^{\uparrow\downarrow}] = 0,
\end{equation}
where $\sigma = \uparrow, \downarrow$.

The real roots are obtained for $\tilde{\gamma}_{xz} \leq \tilde{\gamma}_{xz}^{Max}$, where $\tilde{\gamma}_{xz}^{Max} = \sqrt{\frac{\pi \hbar^4 g_0 \mu_B}{8 m_e^2 a^5 M_S}\frac{\mathrm{Im}[g_{eff}^{\uparrow\downarrow}]^2}{\mathrm{Re}[g_{eff}^{\uparrow\downarrow}]}}$.
It is important to emphasize at this point, that $\mathrm{Im}[g_{eff}^{\uparrow\downarrow}]\neq 0$ gives rise to a finite value of the interface spin-orbit parameter $\tilde{\gamma}_{xz}$.
In particular, for $\tilde{\gamma}_{xz} = \tilde{\gamma}_{xz}^{Max}$, the hopping amplitudes have equal values $\tilde{t}_{\uparrow}^0 = \tilde{t}_{\downarrow}^0 = \sqrt{\frac{\pi \hbar^4 g_0 \mu_B}{2 m_e^2 a^5 M_S}\mathrm{Re}[g_{eff}^{\uparrow\downarrow}]}$.

\begin{figure}[]
 \includegraphics[width=0.9\linewidth]{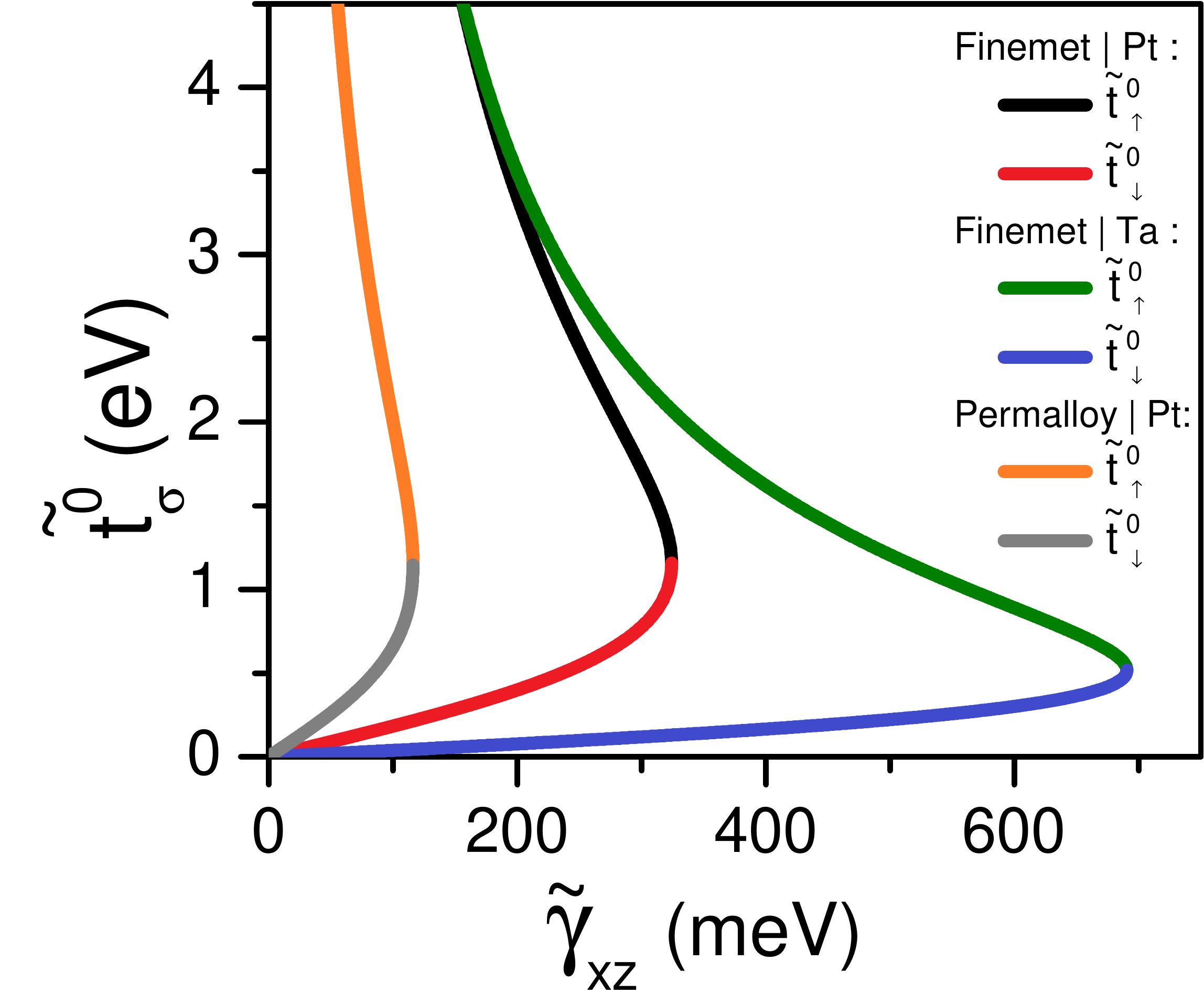}
\caption{Hopping amplitudes $\tilde{t}_{\sigma}^0$ versus interface spin-orbit interaction $\tilde{\gamma}_{xz}$ of Finemet|Pt, Finemet|Ta bilayers.
For comparison, the curves for Permalloy|Pt bilayer were plotted according to parameters found in Ref. \cite{7}.
The lattice constant was taken after Ref. \cite{37} (3.5 \AA).}
\label{fig:6}
\end{figure}

For experimentally derived parameters of the studied bilayers, the possible solutions of Eq. \ref{eq:13} are plotted in Fig. \ref{fig:6}.
It is clearly seen that even for a small deviation of $\tilde{\gamma}_{xz}$ from $\tilde{\gamma}_{xz}^{Max}$ the ratio $\tilde{t}_{\uparrow}^0/\tilde{t}_{\downarrow}^0$ experiences a serious increase.
In the case of Finemet|Pt for instance, $\tilde{t}_{\uparrow}^0 = 4.4$ eV, $\tilde{t}_{\downarrow}^0 = 0.32$ eV at $\tilde{\gamma}_{xz}^{Max}/2$, and the ratio $\tilde{t}_{\uparrow}^0/\tilde{t}_{\downarrow}^0 = 13.2$.
We argue that such a result is unphysical, and therefore conclude that the value of interface spin-orbit interaction lies in a close vicinity to $\tilde{\gamma}_{xz}^{Max}$, where $\tilde{t}_{\uparrow}^0 \approx \tilde{t}_{\downarrow}^0$.
For the sake of comparison between different bilayers, we claim that it is valid to consider that $\tilde{\gamma}_{xz} \cong \tilde{\gamma}_{xz}^{Max}$.

For Finemet|Pt $\tilde{\gamma}_{xz}^{Max} = 330$ meV, while for the bilayer with Ta is roughly two times larger (680 meV).
A comparison with Permalloy|Pt \cite{7}, for which $\tilde{\gamma}_{xz}^{Max} = 110$ meV, suggests, that the Finemet based bilayers exhibit distinctively stronger interface spin-orbit interaction.
In contrast, the values of hopping amplitudes lies in the same range for both Permalloy|Pt and Finemet|Pt systems ($\approx 1.16$ eV), although for Finemet|Ta bilayer they are noticeably lower in value ($\approx 0.51$ eV).
The origin of those differences requires further first-principle studies.

\begin{figure*}[]
 \includegraphics[width=0.7\linewidth]{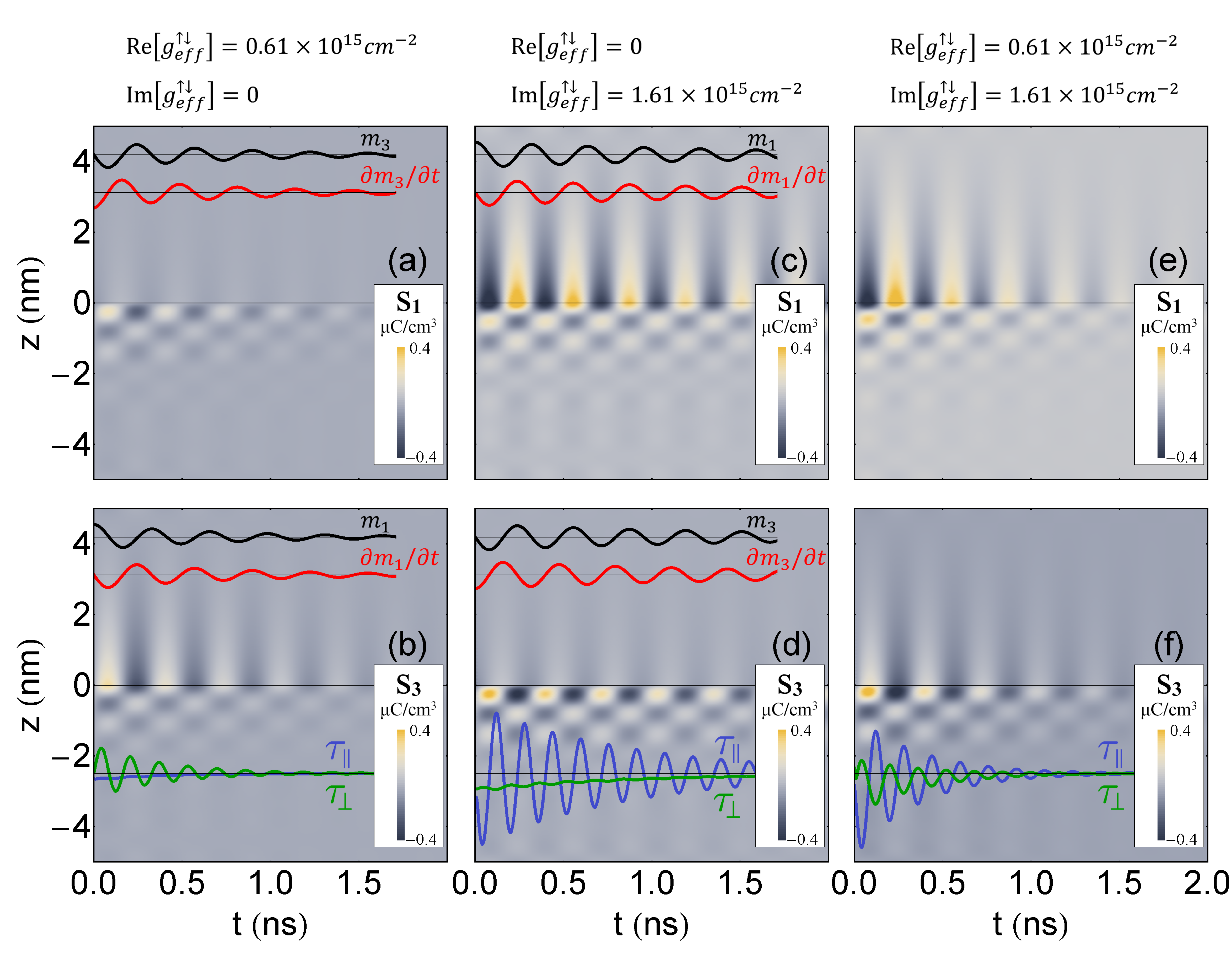}
\caption{The coupled spin-magnetization dynamics in the Finemet|Ta bilayer.
Temporal and spatial distributions of the components $\vec{s}$ ($s_1$ - upper panel and $s_3$ - bottom panel; yellow/grey maps) compared with the temporal evolution of $\vec{m}$ (black lines) and its derivative (red lines) as a consequence of the spin pumping driven by: the real part ((a) and (b)), imaginary part ((c) and (d)) and total ((e) and (f)) spin mixing conductance.
The amplitudes of $\vec{m}$ components were normalized for clarity.
Blue and green lines in (b), (d) and (f) shows value of the spin-transfer torque parallel and perpendicular to the dynamic components of magnetization.
In the adopted reference system, $z = 0$ nm corresponds to the Finemet|Ta interface.
The area below it ($z < 0$) shows the Finemet film and the area above ($z > 0$), the Ta film.}
\label{fig:7}
\end{figure*}

\subsection{Impact of the imaginary part of spin mixing conductance on magnetization dynamics}
\label{IofIPoSMCoMD}

Since the Finemet|Ta exhibits distinctively different properties from Finemet|Pt bilayers, namely the imaginary part of spin mixing conductance of the former is higher than the real part, using the diffusive model, we investigated the spin-magnetization dynamics in this particular system.
The coupled differential equations for the magnetization $\vec{m}$, spin accumulation $\vec{s}$ and spin current $\vec{J}_s$ with spin pumping at Finemet|Ta interface and spin torques (STs) terms were solved numerically in time-dependent Comsol Multiphysics model (see Supplementary Materials for detailed information \cite{supp}).
The magnetization was deviated from the equilibrium direction along $\hat{x}_2$ axis to $\vec{m}_0$ = (10 A/m, $M_s$, 0) at $t = 0$ and its precession was being subject to the Gilbert damping and ST.

Due to the spin pumping, an additional spin flux [Eq. (S15)] results in the flow of the non-equilibrium spin density $\vec{s}$ away from Finemet|Ta interface.
In Finemet, the spin accumulation interacts with the magnetization through ST.
In the case of a thin Ta film, i.e., if $d_{Ta}\ll \lambda_{sf}$, the spin pumping flow is immediately balanced by the spin accumulation gradient at the opposite Ta surface and the overall influence on the magnetization dynamics is reduced.
Otherwise, for a thick Ta film (as used in the experimentally investigated samples), the spin relaxation in Ta leads to the loss of total angular momentum in the system and consequently, the damping and the frequency shift of magnetization dynamics is enhanced in Finemet.

Figure \ref{fig:7} shows a temporal evolution of spin accumulation $\vec{s}$ through a thickness of Ta and Finemet (yellow/grey maps).
The real part of spin mixing conductance $\mathrm{Re}[g_{eff}^{\uparrow\downarrow}]$ produces a spin current and a spin accumulation at the F|Ta interface: $(s_1, 0, s_3) \propto (\frac{\delta m_3}{\delta t}, 0, \frac{\delta m_1}{\delta t})$, as shown in Figs. \ref{fig:7}(a) and \ref{fig:7}(b).
The imaginary part of spin mixing conductance $\mathrm{Im}[g_{eff}^{\uparrow\downarrow}]$ on the other hand, produces spin current and spin accumulation at the F|Ta interface: $(s_1, 0, s_3) \propto (\frac{\delta m_1}{\delta t}, 0, \frac{\delta m_3}{\delta t})$, as shown in Figs. \ref{fig:7}(c) and \ref{fig:7}(d).
The difference in amplitudes between $s_1$ and $s_3$ results from the ellipticity of magnetization precession and the higher value of the in-plane component $m_1$ in relation to $m_3$.
It is clearly seen that for the system with Ta, the spin accumulation is governed by $\mathrm{Im}[g_{eff}^{\uparrow\downarrow}]$, which arises from the ratio $\frac{\mathrm{Re}[g_{eff}^{\uparrow\downarrow}]}{\mathrm{Im}[g_{eff}^{\uparrow\downarrow}]} = 0.38$.
Figures \ref{fig:7}(e) and \ref{fig:7}(f) shows the overall dynamics of $\vec{s}$ due to both $\mathrm{Im}[g_{eff}^{\uparrow\downarrow}]$ and $\mathrm{Re}[g_{eff}^{\uparrow\downarrow}]$, that can be also viewed as a resultant of the amplitudes shown in (a), (c) and (b), (d), respectively.

At a given time, spin accumulation decays exponentially in Ta by the factor of $1/e$ at the distance $\lambda_{sf}$ away from the interface. In Finemet, $\vec{s}$ rotates around $\vec{m}$ along the direction $-\hat{x}_3$ from the F|Ta interface (with the characteristic length $2\pi\lambda_{L}=1.8$ nm) leading to the oscillatory behavior of $\vec{s}$ components across Finemet thickness. Since we assumed a linear regime of magnetization dynamics, $\vec{m}\approx M_s \hat{x}_2$, then $\vec{s}$ is always perpendicular to $\vec{m}$ and decays with the characteristic length $\lambda_\bot=1.2$ nm.

To elucidate the influence of $\mathrm{Re}[g_{eff}^{\uparrow\downarrow}]$ and $\mathrm{Im}[g_{eff}^{\uparrow\downarrow}]$ on the temporal evolution of the magnetization we calculated the spin torque in Finemet that is parallel to the dynamic magnetization ($\vec{\tau}_{\parallel}$) and thus it is responsible for the damping and ST perpendicular to the dynamic magnetization ($\vec{\tau}_{\bot}$), responsible for a change of the frequency of the precession (see Supplementary Materials \cite{supp}).
The values of STs have been integrated over the thickness of Finemet and plotted in Fig. \ref{fig:7}.
The magnitude of the perpendicular component of ST $\tau_{\bot}$, generated by the real part of the spin mixing conductance (green line in Fig. \ref{fig:7}b), oscillates with the frequency that is double of the FMR frequency.
Therefore, the time-averaged value of $\tau_{\bot}$ is zero and the frequency of precession of $\vec{m}$ is not affected.
On the other hand, the value of the parallel component of ST $\tau_{\parallel}$ (blue line in Fig. \ref{fig:7}b) has the negative sign at any time.
This means that ST is always antiparallel to the dynamic components of magnetization, leading to the enhanced damping.
For the imaginary part of spin mixing conductance it is $\tau_{\parallel}$ which oscillates with the frequency that is double of the FMR frequency (blue line in Fig. \ref{fig:7}(d)) resulting in almost zero damping.
The value of the perpendicular component of ST $\tau_{\bot}$ (green line in Fig. \ref{fig:7}(d)) has the negative sign at any time, acting always to increase the frequency of precession.
In Fig. \ref{fig:7}(f) the overall STs are shown, which are the sum of the influence of real and imaginary part of the spin mixing conductance.

\section{CONCLUSIONS}
\label{CS}
In summary, we showed that the investigated spin pumping effect in Finemet films with Pt and Ta capping layers results in linear dependences of the Gilbert damping parameter and the $g$-factor shift on the inverse Finemet thickness what subsequently allowed us to determine both real and imaginary part of the spin mixing conductance.
Giving the insight into microscopic origin of this behavior, we demonstrated in particular, that the magnetization dynamics in Finemet|Ta bilayers is governed by a dominant role of $\mathrm{Im}[g_{eff}^{\uparrow\downarrow}]$, which arises from the strong interface spin-orbit interaction and contributes to the field-like torque.
This turns out to be uncommon but from an application point of view, this fact offers an attractive perspective in the field of spin-orbitronics.
A small enhancement of the precession damping combined with relatively large torque exerted on the magnetization and low saturation magnetization of the ferromagnetic layer indicate the possibility of efficient magnetization switching in next-generation magnetic random-access memory cells.

\section*{Acknowledgements}
This work was supported by the project “Marie Skłodowska-Curie Research and Innovation Staff Exchange (RISE)” Contract No. 644348 with the European Commission, as part of the Horizon2020 Programme. A. K. acknowledges the support from program POWR.03.02.00-00-I032/16.

\bibliography{references}

\end{document}